# The effective Hamiltonian which governs the propagation dynamics of nonspreading wave packets

Chyi-Lung Lin,
Department of Physics, Soochow University,
Taipei 111, Taiwan, R.O.C.

**ABSTRACT**

We discuss the propagation dynamics of nonspreading wave packets $\Psi(x, t)$. We decompose the Hamiltonian H into two parts: $H = \widetilde{H}(t) + H_c(t)$. The first part $\widetilde{H}(t)$ is such that $\Psi(x, t)$ is its instantaneous eigenstate and is therefore irrelevant to the propagation of the packet. The second part $H_c(t)$ is shown to be the effective Hamiltonian governing the motion of the packet both classically and quantum mechanically. Thus, analogous to Ehrenfest's theorem, nonspreading wave packets offer another view point directly connecting quantum mechanics and classical mechanics. This analysis also works for non-square-integrable packets, such as Airy packets.

E-mail: cllin@scu.edu.tw

# I. Introduction

In a recent paper, Noa Voloch-Bloch et al. showed the generation of nonspreading electron Airy beams [1]. This is a further exciting development since the first observation of optical Airy beams by Siviloglou et al. in 2007 [2]. One of the reasons in constructing Airy beams is because wave packets based on Airy functions remain nonspreading during free propagation. That is we need not prepare any particular environment for Airy packets to be nonspreading; just simply a free space. However, we should note that what the experiments have constructed are actually truncated Airy beams; this is due to that an ideal Airy packet is with infinite energy. Also, the constructed Airy beams are in spatial dimensions. We here call attention to the time evolution of nonspreading wave packets. In what follows, we shall denote nonspreading wave packet by NWP.

In 1926, Schrodinger constructed the first NWP. The profile being used was taken from that of the ground state of simple harmonic oscillator (abbreviated as SHO) [3]. After that, in 1954 Senitzky generalized Schrodinger's result to constructing NWPs using the shapes of higher energy eigenstates of SHO [4]. Both results show that the motion of these NWPs is the same as that of a classical particle in SHO. This is consistent with Ehrenfest's theorem [5-6]. However, in 1979 Berry and Balazs found a NWP in free space [7]. This NWP is with the shape of an Airy function. Surprisingly, it moves with a constant acceleration without any external driving field. This shows that a quantum Airy packet doesn't move like a classical particle in the case of free space. This strange phenomenon, however, does not contradict with Ehrenfest's theorem, as an Airy packet is not square-integrable, hence, expectation values cannot be defined. In consequence, Ehrenfest's theorem does not describe the corresponding classical behavior for non-square-integrable packets. We will show below that it is possible to understand the propagation dynamics of NWPs from a suitable Hamiltonian.

The main point is that the Hamiltonian $\mathbf{H}$ although governs the time evolution of a wave packet; however, the Hamiltonian governing the *motion* of NWP should not be read directly from the original Hamiltonian $\mathbf{H}$. As in quantum mechanics, a wave packet is stationary (motionless) only when it is the energy eigenstate of a Hamiltonian.

Suppose we have a packet which is a stationary state of a Hamiltonian, denoted by $\widetilde{\mathbf{H}}$. We next consider a Hamiltonian $\mathbf{H}$ which contains an extra term $\mathbf{H_c}$ besides $\widetilde{\mathbf{H}}$. The packet now is not stationary, and due to that $\widetilde{\mathbf{H}}$ is irrelevant to the motion of a packet, it should be $\mathbf{H_c}$ that governs the *change* or the *motion* of the packet.

With this view point, when considering the motion of a NWP $\mathbf{\Psi(x,t)}$, we should first find out the Hamiltonian $\widetilde{\mathbf{H}}$, such that $\mathbf{\Psi(x,t)}$ is its instantaneous eigenstate. In general, $\widetilde{\mathbf{H}}$ is time dependent, hence we denote it by $\widetilde{\mathbf{H}}(\mathbf{t})$. We have the following eigen-equation of $\widetilde{\mathbf{H}}$:

$$\widetilde{\mathbf{H}}(\mathbf{t})\, \mathbf{\Psi(x,t)} = \tilde{\mathbf{E}}(\mathbf{t})\, \mathbf{\Psi(x,t)} \qquad (1)$$

The eigenvalue of $\widetilde{\mathbf{H}}(\mathbf{t})$ is denoted by $\tilde{\mathbf{E}}(\mathbf{t})$**.** We then decompose the original Hamiltonian $\mathbf{H}$ into

$$\mathbf{H} = \widetilde{\mathbf{H}}(\mathbf{t}) + \mathbf{H_c(t)} \qquad (2)$$

It should be $\mathbf{H_c(t)}$ governing the motion of the packet. This can be proved from the infinitesimal time evolution operator $\mathbf{U(t+dt,t)}$, which is defined as

$$\mathbf{U(t+dt,t)} = \mathbf{exp}\left[-\mathbf{i}\,\frac{\mathbf{H}}{\hbar}\mathbf{dt}\right] \qquad (3)$$

Then

$$\mathbf{\Psi(x,t+dt)} = \mathbf{U(t+dt,t)}\,\mathbf{\Psi(x,t)} \qquad (4)$$

Using the decomposition formula (2) for $\mathbf{H}$, $\mathbf{U(t+dt,t)}$ then contains two terms. One is the operator $\mathbf{exp}\left[-\mathbf{i}\,\frac{\widetilde{\mathbf{H}}(\mathbf{t})}{\hbar}\mathbf{dt}\right]$; when acting on $\mathbf{\Psi(x,t)}$**,** this term only gives a phase factor $\mathbf{exp}\left[-\mathbf{i}\,\frac{\tilde{\mathbf{E}}(\mathbf{t})}{\hbar}\mathbf{dt}\right]$**.** Hence $\widetilde{\mathbf{H}}(\mathbf{t})$ does not affect the state of the motion and therefore does not relate to the propagation**.** The other is the operator $\mathbf{exp}\left[-\mathbf{i}\,\frac{\mathbf{H_c(t)}}{\hbar}\mathbf{dt}\right]$**.** We then have

$$\mathbf{U(t+dt,t)}\,\mathbf{\Psi(x,t)}$$

$$= \mathbf{exp}\left[-\mathbf{i}\ \frac{\tilde{\mathbf{E}}(\mathbf{t})}{\hbar}\mathbf{dt}\right]\ \mathbf{exp}\left[-\mathbf{i}\ \frac{\mathbf{H_c}(\mathbf{t})}{\hbar}\mathbf{dt}\right]\boldsymbol{\Psi}(\mathbf{x},\mathbf{t}) \qquad (5)$$

If $\boldsymbol{\Psi}(\mathbf{x},\mathbf{t})$ is an arbitrary packet, then $\mathbf{exp}\left[-\mathbf{i}\ \frac{\mathbf{H_c}(\mathbf{t})}{\hbar}\mathbf{dt}\right]$ will cause the change of the shape. However, if $\boldsymbol{\Psi}(\mathbf{x},\mathbf{t})$ is a NWP, then the action of $\mathbf{exp}\left[-\mathbf{i}\ \frac{\mathbf{H_c}(\mathbf{t})}{\hbar}\mathbf{dt}\right]$ will not change the shape but only make a spatial shift to the packet. The motion of a NWP is then determined by $\mathbf{H_c}(\mathbf{t})$ through the operator $\mathbf{exp}\left[-\mathbf{i}\ \frac{\mathbf{H_c}(\mathbf{t})}{\hbar}\mathbf{dt}\right]$. Thus, from the point of view of quantum mechanics, it is effectively the Hamiltonian $\mathbf{H_c}(\mathbf{t})$ that governs the motion of NWPs.

As $\widetilde{\mathbf{H}}(\mathbf{t})$ is not related to motion, and since it is intuitive to model a NWP as a classical particle, we expect that the motion of this "particle" is governed by $\mathbf{H_c}(\mathbf{t})$ in the usual classical Hamiltonian formalism. In classical mechanics, the dynamics from the Hamiltonian $\mathbf{H_c}(\mathbf{t})$ is

$$\dot{\mathbf{x}} = \frac{\partial\, \mathbf{H_c}}{\partial\, \mathbf{p}} \qquad (6)$$

$$\dot{\mathbf{p}} = -\frac{\partial\, \mathbf{H_c}}{\partial\, \mathbf{x}} \qquad (7)$$

We then expect that the motion described in (6-7) is the same as that of the NWP. We will show below that this is true. Thus, NWPs offer another view point besides Ehrenfest's theorem connecting quantum mechanics and classical mechanics.

We conclude with the following statement:

Although the Hamiltonian $\mathbf{H}$ governs the time evolution of a wave packet; however, it is $\mathbf{H_c}(\mathbf{t})$ which governs the propagation of a NWP both quantum mechanically and classically**.**

In what follows, we examine this statement for the following three cases of NWPs.

## (I) NWP in free space

In free space, the Hamiltonian is $\mathbf{H} = \mathbf{H_0} = \frac{\mathbf{p^2}}{\mathbf{2m}} = -\frac{\hbar^2}{\mathbf{2m}}\frac{\partial^2}{\partial \mathbf{x^2}}$. From the result of Berry and Balazs [7], the NWP in free space is described by an Airy function $\mathbf{Ai[x,t]}$. We have

$$\boldsymbol{\Psi}(\mathbf{x},\mathbf{t}) = \mathbf{Ai}\,[\mathbf{b}\,(\mathbf{x} - \mathbf{d_0}(\mathbf{t}))]\,\mathbf{exp}\left[\frac{\mathbf{i}}{\hbar}\boldsymbol{\phi}\,(\mathbf{x},\mathbf{t})\right] \tag{8}$$

where b is an arbitrary constant, and

$$\mathbf{d_0}(\mathbf{t}) = \frac{\mathbf{f_b t^2}}{\mathbf{2m}}, \qquad \mathbf{f_b} \equiv \frac{\hbar^2\mathbf{b^3}}{\mathbf{2m}} \tag{9}$$

$$\boldsymbol{\phi}(\mathbf{x},\mathbf{t}) = \mathbf{m}\,\dot{\mathbf{d}}_\mathbf{0}(\mathbf{t})\,\mathbf{x} + \boldsymbol{\phi}_\mathbf{0}(\mathbf{t}), \quad \boldsymbol{\phi}_\mathbf{0}(\mathbf{t}) = -\frac{\mathbf{f_b}^2\,\mathbf{t^3}}{\mathbf{3m}} \tag{10}$$

Obviously, the packet $\boldsymbol{\Psi}(\mathbf{x},\mathbf{t})$ is nonspreading and moves with a trajectory $\mathbf{x} = \mathbf{d_0}(\mathbf{t})$, which is a motion with a constant acceleration $\mathbf{a} = \mathbf{f_b}/\mathbf{m}$. *To understand the appearance of this acceleration in free space*, we first know that an Airy function of the form as $\mathbf{Ai}[\mathbf{b}\,(\mathbf{x} - \mathbf{d}))]$ is an eigenfunction of the Hamiltonian $\mathbf{H_b}$, defined as:

$$\mathbf{H_b} \equiv \mathbf{H_0} + \mathbf{f_b}\;\mathbf{x} \tag{11}$$

Then

$$\mathbf{H_b}\;\mathbf{Ai}[\mathbf{b}(\mathbf{x} - \mathbf{d})] = \mathbf{E_b}\;\mathbf{Ai}[\mathbf{b}(\mathbf{x} - \mathbf{d})] \tag{12}$$

$$\mathbf{E_b} = \mathbf{f_b}\,\mathbf{d} \tag{13}$$

We next find the Hamiltonian $\widetilde{\mathbf{H}}(\mathbf{t})$ such that $\boldsymbol{\Psi}(\mathbf{x},\mathbf{t})$ is its eigenstate. With $\boldsymbol{\Psi}(\mathbf{x},\mathbf{t})$ given by (8), we easily find

$$\widetilde{\mathbf{H}}(\mathbf{t}) = -\frac{\hbar^2}{\mathbf{2m}}\frac{\partial^2}{\partial \mathbf{x^2}} - \dot{\mathbf{d}}_\mathbf{0}(\mathbf{t})\mathbf{p} + \mathbf{f_b}\,\mathbf{x} \tag{14}$$

$$\widetilde{\mathbf{E}}(\mathbf{t}) = \mathbf{f_b}\,\mathbf{d_0}(\mathbf{t}) - \frac{\mathbf{1}}{\mathbf{2}}\mathbf{m}\,\dot{\mathbf{d}}_\mathbf{0}(\mathbf{t})^\mathbf{2} \tag{15}$$

where $\dot{\mathbf{d}}_0(\mathbf{t})$ means the time derivative of $\mathbf{d}_0(\mathbf{t})$**.** Using (9), we see that in fact $\tilde{\mathbf{E}}(\mathbf{t}) = \mathbf{0}$**.** We now decompose the original Hamiltonian $\mathbf{H}_0$ into

$$\mathbf{H}_0 = \widetilde{\mathbf{H}}(\mathbf{t}) + \mathbf{H}_c(\mathbf{t}) \tag{16}$$

$$\mathbf{H}_c(\mathbf{t}) = \dot{\mathbf{d}}_0(\mathbf{t})\,\mathbf{p} - \mathbf{f}_b\,\mathbf{x} \tag{17}$$

Firstly, we show that $\mathbf{H}_c$ governs the propagation dynamics of NWP quantum mechanically. From (17), we have

$$\mathbf{exp}\left[-\mathbf{i}\,\frac{\mathbf{H}_c(\mathbf{t})}{\hbar}\mathbf{dt}\right] = \mathbf{exp}\left[\mathbf{i}\,\frac{\mathbf{f}_b\,\mathbf{x}}{\hbar}\mathbf{dt}\right]\,\mathbf{exp}\left[-\mathbf{i}\,\frac{\dot{\mathbf{d}}_0(\mathbf{t})\mathbf{p}}{\hbar}\mathbf{dt}\right] \tag{18}$$

Substitute (18) to (5), we have

$$\begin{aligned}&\mathbf{U}(\mathbf{t}+\mathbf{dt},\mathbf{t})\;\boldsymbol{\Psi}(\mathbf{x},\mathbf{t})\\ &= \mathbf{exp}\left[\mathbf{i}\,\frac{\mathbf{f}_b\,\mathbf{x}-\tilde{\mathbf{E}}(\mathbf{t})}{\hbar}\mathbf{dt}\right]\,\mathbf{exp}\left[-\mathbf{i}\,\frac{\dot{\mathbf{d}}_0(\mathbf{t})\mathbf{p}}{\hbar}\mathbf{dt}\right]\;\boldsymbol{\Psi}(\mathbf{x},\mathbf{t})\end{aligned} \tag{19}$$

We see that it is the spatial shift operator $\mathbf{exp}\left[-\mathbf{i}\,\frac{\dot{\mathbf{d}}_0(\mathbf{t})\,\mathbf{p}}{\hbar}\mathbf{dt}\right]$ that causes the propagation of NWP, and hence the acceleration. When acting on $\boldsymbol{\Psi}(\mathbf{x},\mathbf{t})$, it makes a spatial shift to the packet with the amount $\mathbf{dx} = \dot{\mathbf{d}}_0(\mathbf{t})\mathbf{dt}$. The packet then moves with a velocity $\mathbf{v} = \dot{\mathbf{d}}_0(\mathbf{t}) = \frac{\mathbf{f}_b\mathbf{t}}{\mathbf{m}}$**,** or with an acceleration $\mathbf{a} = \ddot{\mathbf{d}}_0(\mathbf{t}) = \mathbf{f}_b/\mathbf{m}$. Thus, in *free space*, through the specific form of $\mathbf{H}_c(\mathbf{t})$ in (17), an Airy packet is pushed into a constant acceleration motion. Hence, it is $\mathbf{H}_c(\mathbf{t})$ that governs the propagation of a NWP. This can be generalized to study how $\mathbf{H}_c(\mathbf{t})$ governs the *change* of any arbitrary wave function.

Secondly, we study the classical motion derived from $\mathbf{H}_c$. In classical mechanics, the dynamics described by $\mathbf{H}_c(\mathbf{t})$ is:

$$\dot{\mathbf{x}} = \frac{\partial \mathbf{H}_c}{\partial \mathbf{p}} = \dot{\mathbf{d}}_0(\mathbf{t}) \tag{20}$$

$$\dot{\mathbf{p}} = -\frac{\partial \mathbf{H}_c}{\partial \mathbf{x}} = \mathbf{f}_b \tag{21}$$

Formula (20) describes the particle moving with the trajectory $\mathbf{x} = \mathbf{d_0}(\mathbf{t})$. Thus the classical motion derived from $\mathbf{H_c}(\mathbf{t})$ is the same as the propagation of a NWP. A NWP can indeed be treated as a classical particle whose motion obeys the laws of classical mechanics with $\mathbf{H_c}(\mathbf{t})$ as the Hamiltonian. We also note that formula (21) shows that the classical force law is $\dot{\mathbf{p}} = \mathbf{f_b} = \mathbf{m}\,\ddot{\mathbf{d}}_\mathbf{0}(\mathbf{t}) = \mathbf{m}\,\ddot{\mathbf{x}}$.

We may also discuss the finite time evolution operator $\mathbf{U}(\mathbf{t},\mathbf{0})$. From which, we can obtain $\mathbf{\Psi}(\mathbf{x},\mathbf{t})$ from an initial wave packet $\mathbf{\Psi}(\mathbf{x},\mathbf{0})$. One of the methods for obtaining $\mathbf{U}(\mathbf{t},\mathbf{0})$ can be referred to reference [8]. However, we can obtain $\mathbf{U}(\mathbf{t},\mathbf{0})$ from the product of the infinitesimal time evolution operator at each instant of time. That is we use the formula

$$\mathbf{U}(\mathbf{t},\mathbf{0})\mathbf{\Psi}(\mathbf{x},\mathbf{0}) = \prod_{\mathbf{i}=\mathbf{0}}^{\mathbf{N}-\mathbf{1}} \mathbf{U}(\mathbf{t_{i+1}},\mathbf{t_i})\,\mathbf{\Psi}(\mathbf{x},\mathbf{0}) \qquad (22)$$

where $\mathbf{t_i} = \mathbf{i}\,\Delta\mathbf{t}$, and $\Delta\mathbf{t} = \mathbf{t}/\mathbf{N} \to \mathbf{0}$. Using the result of (19) for each $\mathbf{U}(\mathbf{t_{i+1}},\mathbf{t_i})$**,** and after some calculation, we obtain

$$\begin{aligned}&\mathbf{U}(\mathbf{t},\mathbf{0})\,\mathbf{\Psi}(\mathbf{x},\mathbf{0})\\ &= \mathbf{exp}\left[\frac{\mathbf{i}}{\hbar}\,\boldsymbol{\phi}(\mathbf{x},\mathbf{t})\right]\,\mathbf{exp}\left[\frac{-\mathbf{i}}{\hbar}\,\mathbf{d_0}(\mathbf{t})\,\mathbf{p}\right]\,\mathbf{\Psi}(\mathbf{x},\mathbf{0})\end{aligned} \qquad (23)$$

where $\boldsymbol{\phi}(\mathbf{x},\mathbf{t})$ and $\mathbf{d_0}(\mathbf{t})$ are defined in (9-10). Formula (23) shows that the packet is nonspreading and moves with a trajectory $\mathbf{x} = \mathbf{d_0}(\mathbf{t})$. With $\mathbf{\Psi}(\mathbf{x},\mathbf{0}) = \mathbf{Ai}[\mathbf{b}\,\mathbf{x}]$, then from (23) we recover the $\mathbf{\Psi}(\mathbf{x},\mathbf{t})$ in formula (8).

**(II) NWP in a time-varying spatially uniform linear potential**

We next discuss NWPs in a system with a time dependent Hamiltonian $\mathbf{H}(\mathbf{t})$, which is defined as

$$\mathbf{H}(\mathbf{t}) = \mathbf{H_0} - \mathbf{F}(\mathbf{t})\,\mathbf{x} \qquad (24)$$

We first define

$$\boldsymbol{\alpha}(\mathbf{t}) = \int_\mathbf{0}^\mathbf{t} \mathbf{F}(\boldsymbol{\tau})\,\mathbf{d}\boldsymbol{\tau} \qquad (25)$$

From Berry and Balazs [7], the wave function of the NWP is:

$$\mathbf{\Psi(x,t) = Ai\ [b\ (x - d(t))]\ \ exp\left[\frac{i}{\hbar}\phi(x,t)\right]} \qquad (26)$$

where

$$\mathbf{d(t) = d_0(t) + d_1(t)} \qquad (27)$$

The $\mathbf{d_0(t)}$ in (27) is given by (9), and

$$\mathbf{d_1(t) = \int_0^t \frac{\alpha(\tau)}{m}\ d\tau = \int_0^t \int_0^\tau \frac{F(s)}{m} ds\ d\tau} \qquad (28)$$

The trajectory of the motion described by (26) is $\mathbf{x = d(t)}$, which represents the quantum packet moving with an acceleration $\mathbf{a} = \frac{\mathbf{f_b+F(t)}}{\mathbf{m}}$. The total phase $\mathbf{\phi(x,t)}$ in (26) is given by

$$\mathbf{\phi\ (x,t) = m\ \dot{d}(t)\ x + \phi_0(t)} \qquad (29)$$

$$\mathbf{\phi_0(t) = -\frac{{f_b}^2 t^3}{3m} - f_b\ t\ d_1(t) - \frac{1}{2m}\int_0^t\ \alpha(\tau)^2 d\tau} \qquad (30)$$

From the $\mathbf{\Psi(x,t)}$ given in (26), we can find $\mathbf{\widetilde{H}(t)}$ and $\mathbf{\tilde{E}(t)}$ as:

$$\mathbf{\widetilde{H}(t) = -\frac{\hbar^2}{2m}\frac{\partial^2}{\partial x^2} - \dot{d}(t)p + \ f_b\ x} \qquad (31)$$

$$\mathbf{\tilde{E}(t) = f_b\ d(t) - \frac{1}{2}m\ \dot{d}(t)^2} \qquad (32)$$

We then decompose the original Hamiltonian $\mathbf{H(t)}$ into the form as

$$\mathbf{H(t) = \widetilde{H}(t) + H_c(t)} \qquad (33)$$

$$\mathbf{H_c(t) = \ \dot{d}(t)p - (f_b + F(t))\ x} \qquad (34)$$

We note that (34) has the same form as (17). It then shows that the packet moves with a velocity $\mathbf{v} = \dot{\mathbf{d}}(\mathbf{t})$, or with an acceleration $\mathbf{a} = \ddot{\mathbf{d}}(\mathbf{t}) = (\mathbf{f_b} + \mathbf{F}(\mathbf{t}))/\mathbf{m}$.

Also, classically, the dynamics from the Hamiltonian $\mathbf{H_c}(\mathbf{t})$ is

$$\dot{\mathbf{x}} = \frac{\partial \mathbf{H_c}}{\partial \mathbf{p}} = \dot{\mathbf{d}}(\mathbf{t}) \tag{35}$$

$$\dot{\mathbf{p}} = -\frac{\partial \mathbf{H_c}}{\partial \mathbf{x}} = \mathbf{f_b} + \mathbf{F}(\mathbf{t}) \tag{36}$$

Formulas (35-36) show that the particle moves with a trajectory $\mathbf{x} = \mathbf{d}(\mathbf{t})$, and the classical force law is $\dot{\mathbf{p}} = \mathbf{f_b} + \mathbf{F}(\mathbf{t}) = \mathbf{m}\,\ddot{\mathbf{x}}$.

For the finite time evolution operator $\mathbf{U}(\mathbf{t}, \mathbf{0})$, we have from (22) that

$$\begin{aligned} &\mathbf{U}(\mathbf{t}, \mathbf{0})\, \mathbf{\Psi}(\mathbf{x}, \mathbf{0}) \\ &= \mathbf{exp}\left[\frac{\mathbf{i}}{\hbar} \boldsymbol{\phi}\,(\mathbf{x}, \mathbf{t})\right] \mathbf{exp}\left[\frac{-\mathbf{i}}{\hbar} \mathbf{d}(\mathbf{t})\mathbf{p}\right] \mathbf{\Psi}(\mathbf{x}, \mathbf{0}) \end{aligned} \tag{37}$$

where $\mathbf{d}(\mathbf{t})$ and $\boldsymbol{\phi}\,(\mathbf{x}, \mathbf{t})$ are defined in (27) and (29). With $\mathbf{\Psi}(\mathbf{x}, \mathbf{0}) = \mathbf{Ai}[\mathbf{b}\,\mathbf{x}]$, then form (37), we recover the $\mathbf{\Psi}(\mathbf{x}, \mathbf{t})$ in formula (26).

## (III) NWP in a SHO (simple harmonic oscillator)

In this final example, we show that our analysis also applies to square-integrable NWPs. For SHO, we have

$$\mathbf{H} = -\frac{\hbar^2}{2\mathbf{m}} \frac{\partial^2}{\partial \mathbf{x}^2} + \frac{\mathbf{m}}{2} \boldsymbol{\omega}^2 \mathbf{x}^2 \tag{38}$$

From the results of Schrodinger and Senitzky [3-4], the NWPs in SHO are constructed from the well-known wave function $\mathbf{\Psi_n}(\mathbf{x})$, the eigenstate of $\mathbf{H}$ with energy $\mathbf{E_n} = \left(\mathbf{n} + \frac{\mathbf{1}}{\mathbf{2}}\right) \hbar\, \boldsymbol{\omega}$. The NWP in the SHO is given by:

$$\mathbf{\Psi}(\mathbf{x}, \mathbf{t}) = \mathbf{\Psi_n}(\mathbf{x} - \mathbf{d}(\mathbf{t}))\, \mathbf{exp}\left[\frac{\mathbf{i}}{\hbar}\, \boldsymbol{\phi}\,(\mathbf{x}, \mathbf{t})\right] \tag{39}$$

where

$$\mathbf{d}(\mathbf{t}) = \mathbf{A}\cos(\boldsymbol{\omega}\,\mathbf{t} + \boldsymbol{\theta}) \tag{40}$$

$$\boldsymbol{\phi}\,(\mathbf{x},\mathbf{t}) = \mathbf{m}\,\dot{\mathbf{d}}(\mathbf{t})\,\mathbf{x} + \boldsymbol{\phi}_{\mathbf{0}}(\mathbf{t}) \tag{41}$$

$$\boldsymbol{\phi}_{\mathbf{0}}(\mathbf{t}) = -\mathbf{E}_{\mathbf{n}}\,\mathbf{t} - \int_{0}^{t}\left[\frac{\mathbf{m}}{2}\dot{\mathbf{d}}(\boldsymbol{\tau})^{2} - \frac{\mathbf{m}}{2}\boldsymbol{\omega}^{2}\mathbf{d}(\boldsymbol{\tau})^{2}\right]\,\mathbf{d}\boldsymbol{\tau} \tag{42}$$

We note that $\boldsymbol{\Psi}(\mathbf{x},\mathbf{t})$ is square-integrable, and moves with a trajectory $\mathbf{x} = \mathbf{d}(\mathbf{t})$**.** From (39), we find $\widetilde{\mathbf{H}}(\mathbf{t})$ and $\tilde{\mathbf{E}}(\mathbf{t})$ as:

$$\widetilde{\mathbf{H}}(\mathbf{t}) = -\frac{\hbar^{2}}{2\mathbf{m}}\frac{\partial^{2}}{\partial \mathbf{x}^{2}} + \frac{\mathbf{m}}{2}\boldsymbol{\omega}^{2}\mathbf{x}^{2} - \dot{\mathbf{d}}(\mathbf{t})\mathbf{p} - \mathbf{m}\boldsymbol{\omega}^{2}\mathbf{d}(\mathbf{t})\,\mathbf{x} \tag{43}$$

$$\tilde{\mathbf{E}}(\mathbf{t}) = \mathbf{E}_{\mathbf{n}} - \frac{\mathbf{m}}{2}\dot{\mathbf{d}}(\mathbf{t})^{2} - \frac{\mathbf{m}}{2}\boldsymbol{\omega}^{2}\mathbf{d}(\mathbf{t})^{2} \tag{44}$$

We decompose the original Hamiltonian H into the form as

$$\mathbf{H} = \widetilde{\mathbf{H}}(\mathbf{t}) + \mathbf{H}_{\mathbf{c}}(\mathbf{t}) \tag{45}$$

$$\mathbf{H}_{\mathbf{c}}(\mathbf{t}) = \dot{\mathbf{d}}(\mathbf{t})\mathbf{p} + \mathbf{m}\boldsymbol{\omega}^{2}\mathbf{d}(\mathbf{t})\,\mathbf{x} \tag{46}$$

Formula (46) has the same form as (17), it then shows that the packet moves with a velocity $\mathbf{v} = \dot{\mathbf{d}}(\mathbf{t})$; that is with a trajectory $\mathbf{x} = \mathbf{d}(\mathbf{t})$. Thus $\mathbf{H}_{\mathbf{c}}(\mathbf{t})$ governs the propagation dynamics of the NWP. We then recover the results of Schrodinger and Senitzky by the Hamiltonian $\mathbf{H}_{\mathbf{c}}(\mathbf{t})$. On the other hand, classically, the dynamics described by $\mathbf{H}_{\mathbf{c}}(t)$ is

$$\dot{\mathbf{x}} = \frac{\partial \mathbf{H}_{\mathbf{c}}}{\partial \mathbf{p}} = \dot{\mathbf{d}}(\mathbf{t}) \tag{47}$$

$$\dot{\mathbf{p}} = -\frac{\partial \mathbf{H}_{\mathbf{c}}}{\partial \mathbf{x}} = -\mathbf{m}\boldsymbol{\omega}^{2}\mathbf{d}(\mathbf{t}) \tag{48}$$

Formulas (47-48) show that the particle moves with a trajectory $\mathbf{x} = \mathbf{d}(\mathbf{t})$, and the classical force law is $\dot{\mathbf{p}} = -\mathbf{m}\boldsymbol{\omega}^{2}\,\mathbf{x} = \mathbf{m}\ddot{\mathbf{x}}$ .

For the finite time evolution operator, we obtain from (22) that

$$\mathbf{U}(\mathbf{t},\mathbf{0})\ \mathbf{\Psi}(\mathbf{x},\mathbf{0})$$

$$= \mathbf{exp}\left[\frac{\mathbf{i}}{\hbar}\ \widetilde{\boldsymbol{\phi}}\ (\mathbf{x},\mathbf{t})\right]\ \mathbf{exp}\left[\frac{-\mathbf{i}}{\hbar}\big(\mathbf{d}(\mathbf{t})-\mathbf{d}(\mathbf{0})\big)\mathbf{p}\right]\ \mathbf{\Psi}(\mathbf{x},\mathbf{0}) \qquad (49)$$

$$\widetilde{\boldsymbol{\phi}}\ (\mathbf{x},\mathbf{t}) = \mathbf{m}\ \big(\dot{\mathbf{d}}(\mathbf{t})-\dot{\mathbf{d}}(\mathbf{0})\big)\ \mathbf{x} + \boldsymbol{\phi}_{\mathbf{0}}(\mathbf{t}) + \mathbf{m}\ \dot{\mathbf{d}}(\mathbf{0})(\mathbf{d}(\mathbf{t})-\mathbf{d}(\mathbf{0})) \qquad (50)$$

With $\mathbf{\Psi}(\mathbf{x},\mathbf{0}) = \mathbf{\Psi}_{\mathbf{n}}(\mathbf{x}-\mathbf{d}(\mathbf{0}))\ \mathbf{exp}\left[\frac{\mathbf{i}}{\hbar}\ \mathbf{m}\ \dot{\mathbf{d}}(\mathbf{0})\ \mathbf{x}\right]$, then from (49), we recover the $\mathbf{\Psi}(\mathbf{x},\mathbf{t})$ in formula (39).

## (IV) Conclusion

In quantum mechanics, it is the Hamiltonian $\mathbf{H}$ that governs the time evolution of wave functions. For NWPs, from above three examples, we show that the propagation dynamics of NWP is not to be read directly from the Hamiltonian but should be read from the effective Hamiltonian $\mathbf{H}_{\mathbf{c}}(\mathbf{t})$. This is because the Hamiltonian $\widetilde{\mathbf{H}}(\mathbf{t})$ is irrelevant to the motion of NWP. Thus we need to subtract this part of Hamiltonian from the original Hamiltonian **H.** We then show that $\mathbf{H}_{\mathbf{c}}(\mathbf{t})$ governs the propagation dynamics of NWP quantum mechanically. We also show that the classical motion derived from $\mathbf{H}_{\mathbf{c}}(\mathbf{t})$ is the same as the motion of the quantum NWP. This means that NWP can be treated as a classical particle obeying the laws of classical mechanics with Hamiltonian $\mathbf{H}_{\mathbf{c}}(\mathbf{t})$. Thus, NWPs offer another view point connecting quantum mechanics and classical mechanics.

Also, we note that the distinction between $\mathbf{H}$ and $\mathbf{H}_{\mathbf{c}}(\mathbf{t})$ is similar to that between kinetic energy and total energy in special relativity. In which, when discussing the kinetic energy, we should subtract the rest mass energy from the total energy.

We conclude that for a NWP $\mathbf{\Psi}(\mathbf{x},\mathbf{t})$ in a system with Hamiltonian $\mathbf{H}$, the propagation dynamics of $\mathbf{\Psi}(\mathbf{x},\mathbf{t})$, however, is governed effectively by $\mathbf{H}_{\mathbf{c}}(\mathbf{t})$. Or, we may say that alternatively: A NWP in a system with Hamiltonian $\mathbf{H}$ is equivalent to a classical particle in a system with Hamitonian $\mathbf{H}_{\mathbf{c}}(\mathbf{t})$. This is another version of Ehrenfest's theorem in the case of NWPs.

## REFERENCE

1. Noa Voloch-Bloch, Yossi Lereah, Yigal Lilach, Avraham Gover And Ady Arie, “Generation of electron Airy beams”, Nature **494**, 331 (2013)
2. G.A. Siviloglou, J. Broky, A. Dogariu, and D.N. Christodoulides. “Observation of accelerating Airy beams”, Phys. Rev. Lett. **99**, 213901 (2007)
3. E. Schrodinger, Naturwissenften **14**, 664 (1926)
4. I. R. Senitzky, “Harmonic Oscillator wave functions”, The Phyical Review, **95**, 1115 (1954)
5. L. I. Schiff, *Quantum mechanics*, 3ed, McGraw-Hill 1971, p. 5.
6. David J Griffiths, *Introduction to Quantum Mechanics,* second edition ( Pearson Education, Inc. 2005), p. 18.
7. M.V. Berry and N.L. Balazs, “Non spreading wave packets”, Am. J. Phys. **47,** 264 (1979).
8. Gregorio M. A. and de Castro A. S., “A particle moving in a homogeneous time-varying force”, Am. J. Phys. **52**, 557, (1984).